\documentstyle[twoside,fleqn,espcrc2,epsf]{article}

\newcommand{\bee}{\begin{equation}}
\newcommand{\ee}{\end{equation}}
\newcommand{\beea}{\begin{eqnarray}}
\newcommand{\eea}{\end{eqnarray}}
\newcommand{\rme}{{\rm e}}

\newcommand{\AmS}{{\protect\the\textfont2
  A\kern-.1667em\lower.5ex\hbox{M}\kern-.125emS}}

\title{Fixed-point action for fermions in QCD}

\author{T. DeGrand\address{Physics Department, 
        University of Colorado, \\ 
        Boulder, CO 80309 USA}, %
        A. Hasenfratz$^{\rm a}$,
        P. Hasenfratz\address{Institute for Theoretical Physics,
University of Bern, \\
Sidlerstrasse 5, CH-3012 Bern, Switzerland},
        P. Kunszt$^{\rm b}$,
        F. Niedermayer$^{\rm b}$
}
       
\begin{document}

\begin{abstract}
We report our progress constructing a fixed-point action for
fermions interacting with SU(3) gauge fields.
\end{abstract}

\maketitle

\section{INTRODUCTION}
The ultimate goal of any improvement program for QCD is to find
a perfect action with no lattice artifacts: the renormalized trajectory
(RT) of some renormalization group transformation (RGT), parameterized
by  $g^2$ the gauge coupling and $m$ the quark mass.
That is too difficult a task at present, and so we attempt instead
to find the one-loop perfect fixed point (FP) action (at $\beta=\infty$)
and to follow the trajectory in $m$.
Fixed point  (FP) actions\cite{FP} for non-Abelian gauge theories have proven
their ability to improve scaling for physical observables\cite{FP2}
and provide the starting point for the construction of actions for
full QCD.
The outline of our work with fermions is 
similar to that of the MIT group\cite{MIT}
but differs from it in specifics. There is not enough space for us to
describe it in detail, and so we present here only an impressionistic
overview of our progress to date.

We begin with a set of fermionic ($\psi_n$,
$\bar \psi_n$) and gauge field  ($U_\mu(n)$) variables and integrate
them out to construct an action involving coarse-grained
variables $\Psi_{n_b}$, $\bar \Psi_{n_b}$ and $V_\mu(n_b)$
using a renormalization group kernel
\beea
T =  & \beta T_g(U,V) + \nonumber \\
& \kappa \sum_{n_b}(\bar \Psi_{n_b} - \Omega_{n_b,n}\bar\psi_n)
(\Psi_{n_b} - \Omega_{n_b,n}\psi_n) \nonumber \\
\eea
We assume a fine action
\bee
S = \beta S_g(U) + \bar\psi_i \Delta(U)_{ij}\psi_j  .
\ee
$S_g$ and $T_g$ are the action and blocking kernel for the gauge fields.
The renormalization group equation
\bee
\rme^{-S'} = \int d\psi d\bar\psi dU \rme^{-(T+S)}
\ee
has a pure gauge FP at $g^2=0$ ($\beta \rightarrow \infty$). In that limit the
gauge action dominates the integral; its RG equation is given by the same 
steepest-descent equation as for  a pure gauge model
\bee
S^{FP}(V)=\min_{ \{U\} } \left( S^{FP}(U) +T(U,V)\right),
\label{RGG}
\ee
while the fermions sit in the gauge-field background. Their action
remains quadratic in the field variables, and the transformation of
the fermion action is given most easily in terms of the propagator
\beea
(\Delta'(V))^{-1}_{n_b,n'_b} =  & {1\over \kappa} \delta_{n_b,n'_b} \nonumber \\
+  & \Omega(U)_{n_b,n} (\Delta(U))^{-1}_{n,n'}\Omega(U)_{n',n'_b} \nonumber \\
\label{RGF}
\eea
where $U$ and $V$ are related through Eqn. \ref{RGG}.
In all our work we have focussed on a scale factor 2 RG transformation,
for gauge fields the so-called ``Type-1'' transformation of Ref. \cite{FP2}.

\section{FREE MASSLESS FERMIONS}
The formalism for free fermions has been given by Wiese\cite{Wiese}.
We begin with a continuum action for  fermions
which has no doublers and is chirally symmetric.
We  select
a blocking kernel $\Omega$, iterate it to find a fixed point action,
and then tune parameters in $\Omega$ to make the action maximally local.
We have used one in which $\Omega$ is restricted to a hypercube:
$\Omega_{ij}$ is nonzero only if
$j=i\pm\mu$, $i\pm\mu\pm\nu, \dots$ $i\pm\mu\pm\nu\pm\lambda\pm\sigma$.
Each site communicates to $3^4-1=80$ neighbors.
This RGT explicitly breaks chiral symmetry, and so the resulting FP action will
not be chirally invariant. However, the spectrum should be chiral.

 There are many good parameterizations, resulting in
fairly local FP actions.  However,
one ultimately wants to use these actions in simulations, and the
action must be somehow truncated. There are a number of (subjective) criteria
to select a good RGT, based on the properties of the truncated action
(which the RGT does not know about): a good dispersion relation,
$E(p) = |\vec p|$ out to large $|\vec p|$ with no complex roots, good
free-field thermodynamics, $P = 1/3 \sigma T^4$ even at large
discretization, etc.
The``most natural'' truncation of our FP
actions is to a hypercube. 
The free field action then can be parameterized as
\bee
\Delta_0(x) = \lambda(x) + \sum_\mu \gamma_\mu \rho_\mu(x)
\ee
with five nonzero $\lambda$'s and four nonzero $\rho$'s, corresponding
to each of the nonzero offsets.
An example of a dispersion relation for a ``typical'' hypercubic
action is compared to the Wilson action in Figs. 1 and 2.
We show both branches of the hypercubic action's dispersion relation;
all roots are real.
The non-truncated FP action has a perfect dispersion relation $E= |\vec p|$
for all $ \vec p$.

\begin{figure}[htb]
\begin{center}
\vskip -10mm
\leavevmode
\epsfxsize=60mm
\epsfbox[40 50 530 590]{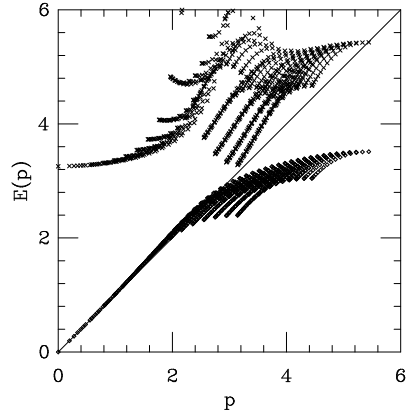}
\vskip -5mm
\end{center}
\caption{Dispersion relation $E(p)$ vs $|p|$ for
 a typical hypercubic approximate FP action.}
\label{fig:disp1}
\end{figure}

\begin{figure}[htb]
\begin{center}
\vskip -10mm
\leavevmode
\epsfxsize=60mm
\epsfbox[40 50 530 590]{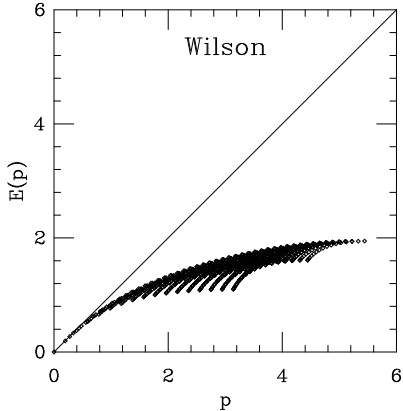}
\vskip -5mm
\end{center}
\caption{Dispersion relation $E(p)$ vs $|p|$ for massless Wilson fermions.}
\label{fig:disp2}
\end{figure}

\section{FREE MASSIVE FERMIONS}
For massive fermions, we want an action which is on an RT
for some RGT (with  $m \rightarrow 2m$ at each step).
 To reach the RT, one can begin with
 an action which has a very small mass
but is otherwise close to a FP action, perform a series of blockings,
and follow it out.

One complication with this procedure is that an action which is local for
small mass can block into an action for large mass which is very nonlocal.
To avoid this, we take an RG transformation whose parameters are functions
of the mass (for example, the parameter $\kappa$ in Eqn. \ref{RGF}
is allowed to vary from step to step) and tune the
 parameter(s) to insure a local
action at each blocking step. The resulting $\lambda$'s and $\rho$'s
are smooth functions of the  mass.
Again, the dispersion relation for hypercubic approximations to
RT actions are well behaved out to large $|\vec p|$.

\section{THE FP VERTEX}
As a first stage in the construction of a FP action for full QCD,
we find the FP vertex: That is, we assume smooth gauge fields,
 expand $U_\mu(x) = 1 + iA_\mu(x)+\dots$, and solve the RG equation
Eq. \ref{RGF}
to lowest nontrivial order in $A_\mu$. The relevant formula has
been given by Bietenholz and Wiese\cite{BW}. 
The fermionic action is
\beea
S = & \sum_{n,n'} \bar \psi(n) \Delta_0(n-n') \psi(n') \nonumber \\
+  & i \sum_{n,n',m}\bar \psi(n) \Delta_1^\mu(n-n',m-n') \nonumber \\
  & \times \psi(n')A_\mu(m). \nonumber \\
\eea
We solve Eq. \ref{RGF} by iteration, beginning with a simple
free action and its associated gauge invariant vertex.
Two independent (but approximate) programs provide cross checks,
and we  have determined the FP vertex to within a few per cent level
for several RGT's.
The vertex $\Delta_1^\mu(n-n',m-n')$ is quite complicated
and contains in addition to scalar and vector terms, many kinds of
``clover'' terms and other operators.  However, most of these
coefficients are very small. The  contributions can be represented
as sums of
paths of $U$'s. For all but the nearest neighbor term, an
(apparently) good approximation to the action
is a simple superposition of all minimum length paths of links
connecting fermions on sites $n$ and $n'$, while for the nearest neighbor
terms, the action involves roughly equal weights of the simple link
$U_\mu(x)$ and a sum of length-3
 paths $U_\nu(x)U_\mu(x+\nu)U^\dagger_\nu(x+\mu)$.
We can also construct the vertex for nonzero mass by iterating the RGT
out along the trajectory as described in the previous section.
The relative weight of  different paths
contributing to the vertex shows little variation
with quark mass while the overall normalization is set by the
parameters of the free action.

\section{NONPERTURBATIVE FP ACTION}
With the perturbative vertex in hand we will construct the full nonperturbative
FP action. This must be done by solving Eqn. \ref{RGF} numerically, then
fitting the action to a set of operators. The perturbative vertex
fixes the operators for smooth gauge fields.
We have not yet done this part of the project, and we do not know yet
whether it is sufficient just to exponentiate the FP vertex, or if
other operators are needed.

We have written various pilot matrix inverters for FP fermion actions.
The cost of matrix inversion, compared to the Wilson action,
is about a factor of twenty per site, which will have to be overcome
by the gain going to a bigger lattice spacing.

We believe the major unsolved problem is to find a parameterization
of the action which is appropriate for coarse
configurations, where we will do simulations.

The final test of an action involves numerical simulation, of course.
We plan to compute spectroscopy at fixed physical volume and
quark mass (fixed $\pi/\rho$ mass ratio) at varying lattice spacing,
without performing any chiral extrapolations.
Our fiducial will be the excellent staggered spectroscopy of the 
MILC collaboration\cite{Sugar}.

This work was supported by the U.S. Department of 
Energy, the U. S. National Science Foundation, and the
Swiss National Science Foundation.


\begin{thebibliography}{9}

\bibitem{FP}
P.~Hasenfratz and F.~Niedermayer, Nucl. Phys. B414 (1994) 785.

\bibitem{FP2}
T.~DeGrand,A.~Hasenfratz, P.~Hasenfratz, F.~Niedermayer,
Nucl. Phys. {B454} (1995) 587;
Nucl. Phys. {B454} (1995) 615;
A. Hasenfratz and T. DeGrand, these proceedings.

\bibitem{MIT}
See the contributions of Bietenholz, Brower, Chandrasekharan, and
Wiese to this volume.

\bibitem{Wiese}
U.~J. Wiese, Phys. Lett. B315 (1993) 417.

\bibitem{BW}
W. Bietenholz and U.~J. Wiese, Nucl. Phys. B464 (1996) 319.

\bibitem{Sugar} The MILC collaboration, in these proceedings.

\end{thebibliography}
\end{document}